\documentclass[prl,twocolumn,superscriptaddress,showpacs]{revtex4}

\usepackage{amsmath}
\usepackage{amssymb}
\usepackage{graphicx}
\usepackage{subfigure}

\newcommand{\be}[1]{\begin{equation}\label{#1}}
\newcommand{\ee}{\end{equation}}
\newcommand{\ba}[1]{\begin{eqnarray}\label{#1}}
\newcommand{\ea}{\end{eqnarray}}

\begin{document}

\title{Destabilization of rotating flows with 
positive shear by azimuthal magnetic fields}

\author{Frank Stefani}
\email{f.stefani@hzdr.de}
\affiliation{Helmholtz-Zentrum Dresden - Rossendorf\\
P.O. Box 510119, D-01314 Dresden, Germany}
\author{Oleg N. Kirillov}
\email{kirillov@mi.ras.ru}
\affiliation{Russian Academy of Sciences, 
Steklov Mathematical Institute\\
Gubkina st. 8, 119991 Moscow, Russia}


\date{\today}

\begin{abstract}

According to Rayleigh's criterion, rotating flows are 
linearly stable when their specific angular momentum 
increases radially outward. The celebrated 
magnetorotational instability opens a way 
to destabilize those flows, as long as the angular
velocity is decreasing outward. 
Using a local approximation 
we demonstrate that even flows with very steep positive 
shear can be destabilized by azimuthal magnetic fields
which are current-free within the fluid. We illustrate
the transition of this instability to a
rotationally enhanced kink-type instability in
case of a homogeneous current in the fluid, and discuss the
prospects for observing it in a magnetized Taylor-Couette flow.

\end{abstract}

\pacs{47.32.-y, 47.35.Tv, 47.85.L-, 97.10.Gz, 95.30.Qd}

\maketitle

From the purely hydrodynamic point of view, rotating flows
are stable as long as their angular momentum is increasing 
radially outward \cite{RAYLEIGH}. Since this criterion 
applies to the Keplerian rotation profiles which are typical 
for low-mass accretion disks, the 
growth mechanism of central objects, such as protostars 
and black holes, had been a conundrum for many decades.
Nowadays, the magnetorotational instability (MRI) 
\cite{VELI}
is considered the main candidate to explain turbulence 
and enhanced angular momentum in accretion disks.
The standard version of MRI (SMRI), with a vertical 
magnetic field $B_z$ applied to the rotating flow,
requires both  the rotation period and the Alfv{\'{e}}n
crossing time to be shorter than the timescale for  
magnetic diffusion  \cite{LIU2006}. This implies,  
for a disk of height
$H$, that both the magnetic Reynolds number 
${\rm Rm}=\mu_0 \sigma H^2 \Omega$ and the Lundquist number 
$S=\mu_0 \sigma H v_{A}$                       
must be larger than one  
($\Omega$ is the angular velocity,
$\mu_0$ is the magnetic permeability constant,
$\sigma$ the conductivity, 
$v_A:=B_z/\sqrt{\mu_0 \rho}$ is the 
Alfv{\'{e}}n velocity, with $\rho$ denoting the density).
While these conditions are safely fulfilled 
in well-conducting parts of accretion disks,
the situation is less clear 
in  the ``dead zones'' of protoplanetary disks, in 
stellar interiors and liquid cores of planets,
because of  the small value of the 
magnetic Prandtl 
number ${\rm Pm}=\nu/\eta$ \cite{BH08},
i.e. the ratio of viscosity $\nu$ to magnetic diffusivity
$\eta:=(\mu_0 \sigma)^{-1}$.

This low  ${\rm Pm}$ case  
is also the subject of intense theoretical and experimental
research initiated by Hollerbach and R\"udiger
\cite{HR95}. Adding an azimuthal magnetic field 
$B_{\phi}$ to $B_{z}$, the authors found a new version of 
MRI, now called helical MRI (HMRI).
It was proved to work also in the inductionless limit \cite{P11},
${\rm Pm}=0$, and to be governed by the Reynolds number
${\rm Re}={\rm Rm} {\rm Pm}^{-1}$ and the Hartmann number
${\rm Ha}=S {\rm Pm}^{-1/2}$,
quite in contrast to standard SMRI
that is governed by ${\rm Rm}$ and $S$.

A somewhat sobering limitation of HMRI was identified 
by Liu et al. \cite{LIU} who used a local approximation
(also called short-wavelength, Wentzel-Kramers-Brillouin (WKB), or 
geometric optics 
approximation, see \cite{KS14}) to
find a minimum steepness of the rotation profile $\Omega(r)$,
expressed by the
Rossby number ${\rm Ro}:= r(2\Omega)^{-1} \partial \Omega/
\partial r$, of 
${\rm Ro}_{\rm LLL}=2(1{-}\sqrt 2)\approx -0.828$.
This {\it lower Liu limit} (LLL) implies that, at least 
for $B_{\phi}(r) \propto 1/r$, HMRI does not 
extend to the
most relevant Keplerian case, characterized by
${\rm Ro}_{\rm Kep}=-3/4$. Surprisingly, in addition to 
the LLL, the authors found also a second threshold of
${\rm Ro}$, which we call
{\it upper Liu limit} (ULL), at
${\rm Ro}_{\rm ULL}=2(1{+}\sqrt 2)\approx +4.828$.
For ${\rm Ro}>{\rm Ro}_{\rm ULL}$ one expects a magnetic 
destabilization of those flows with strongly increasing angular
velocity {\it that would even be stable with respect to SMRI}. 

By relaxing the demand that the azimuthal field is current-free
in the liquid, i.e. $B_{\phi}(r) \propto 1/r$,  and 
allowing fields with arbitrary
radial dependence, we have recently shown \cite{KS14,KS13} that
the LLL and the ULL are just the endpoints of one
common instability curve 
in a plane that is spanned by ${\rm Ro}$ and a 
corresponding steepness of the azimuthal magnetic field, called
magnetic Rossby number, 
${\rm Rb}:=r (2 B_{\phi}/r)^{-1} \partial{(B_{\phi}/r)}/ \partial r$.
In the limit of large ${\rm Re}$ and ${\rm Ha}$, this
curve acquires the closed and simple form 
\begin{equation}
{\rm Rb}=-\frac{1}{8}\frac{({\rm Ro}+2)^2}{{\rm Ro}+1}.
\label{rel}
\end{equation}

A non-axisymmetric ``relative'' of HMRI, the azimuthal MRI (AMRI) 
\cite{TEELUCK}, which appears for purely or dominantly $B_{\phi}$,
has been shown to be governed by basically the same 
scaling behaviour, and the same Liu limits \cite{APJ12}. 
Actually, the key parameter dependencies 
of HMRI and AMRI
were confirmed in various liquid metal experiments 
at the PROMISE facility \cite{PRE,SEIL2}.

In the present paper, we focus exclusively on the 
case of positive ${\rm Ro}$, i.e. on flows whose
{\it angular velocity} (not only the angular frequency) is
increasing outward. From the purely hydrodynamic point of view, such
flows are linearly stable (while non-linear 
instabilities were actually observed in experiments \cite{TSUKAHARA}). 
Flows with positive ${\rm Ro}$ are indeed relevant
for the equator-near strip (approximately between $\pm 30^{\circ}$) 
of the solar tachocline \cite{PAME},
which is, interestingly, also the region of sunspot activity \cite{CHA}.
Up to present, the ULL at ${\rm Ro}_{\rm ULL}=+4.828$ has only been 
predicted in the framework of various local approximations
\cite{LIU,KS13,KS14}, while
attempts to confirm it in a 1-dimensional modal stability 
code on the basis of Taylor-Couette (TC) flows have failed so far 
\cite{RUEPERS}. 
Hence, the questions arise: Is the magnetically triggered flow
instability for ${\rm Ro}>{\rm Ro}_{\rm ULL}$ a real
phenomenon (which would fundamentally modify
the stability criteria for rotating flows in general), or just 
an artifact of the local approximation,
and is there any chance to observe it in a TC experiment?

In order to tackle these problems we restrict our attention here
to non-axisymmetric instabilities, which are the relevant 
ones for pure $B_{\phi}$, and further assume ${\rm Pm}=0$.
Under these assumptions, we had 
recently \cite{KS14} derived the closed 
equation 
\begin{widetext}
\begin{equation}
{\rm Re}^2=\frac{1}{4} \frac{[(1+{\rm Ha}^2n^2)^2-4{\rm Ha}^2{\rm Rb}
(1+{\rm Ha}^2n^2)-4{\rm Ha}^4n^2][1+{\rm Ha}^2(n^2-2{\rm Rb})]^2}
{{\rm Ha}^4{\rm Ro}^2n^2-
[(1+{\rm Ha}^2(n^2-2{\rm Rb}))^2-4{\rm Ha}^4n^2][{\rm Ro}+1]}
\end{equation}
\end{widetext}
for the marginal curves of the instability,
where the following definitions for 
${\rm Re}$, ${\rm Ha}$ and the modified azimuthal wavenumber $n$ 
are used:
\begin{eqnarray}
{\rm Re}&=&\frac{\alpha}{|{\bf k}|^2}\frac{ \Omega(r)}{\nu} \; , \\
{\rm Ha}&=&\frac{\alpha}{|{\bf k}|^2} \frac{B_{\phi}(r)}{r (\mu_0 \rho \eta \nu)^{1/2}} \; ,\\
n&=&m/\alpha \; ,
\end{eqnarray}
with  $\alpha=k_z/|{\bf k}|$ and $|{\bf k}|^2=k^2_r+k^2_z$
defined as functions of the axial and
radial wavenumbers $k_r$ and $k_z$.

Because of its comparably simple form, and the absence of
the ratio $\beta$ of azimuthal to axial magnetic field (which would
play a decisive role for HMRI), Equation (2) allows to easily visualize the
transition from a shear-driven instability of the AMRI-type 
to the current-driven, kink-type
Tayler instability (TI)  \cite{SEIL1}, when going over 
from $\rm Rb=-1$ to $\rm Rb=0$. 

Let us start with the current-free case, $\rm Rb=-1$. Figure 1a shows, 
for varying values of $\rm Ro$ and the
particular case $n=1.4$, the 
marginal curves in the $\rm Ha$-$\rm Re$ plane. 
We see that the critical $\rm Re$ increases
steeply for $\rm Ro$ below 6 which reflects the fact that
we approach $\rm Ro_{\rm ULL}=4.828$ from above. We ask now for
the dominant wavenumbers, as illustrated in
Figure 1b for the particular value ${\rm Ro=5.5}$. 
Evidently, the minimal values of $\rm Re$ and 
$\rm Ha$ (the ``knee'' of the curve) 
appear for $n \sim 1.4$ which represents a rather ``benign''
combination of wavenumbers with $k_r \sim k_z$, so that 
neither the axial nor the radial wavelength of the perturbations
diverges. From this point of view, 
there seems to be no contradiction with the underlying
short-wavelength approximation.

\begin{figure}
    \begin{center}
    \includegraphics[angle=0, width=0.45\textwidth]{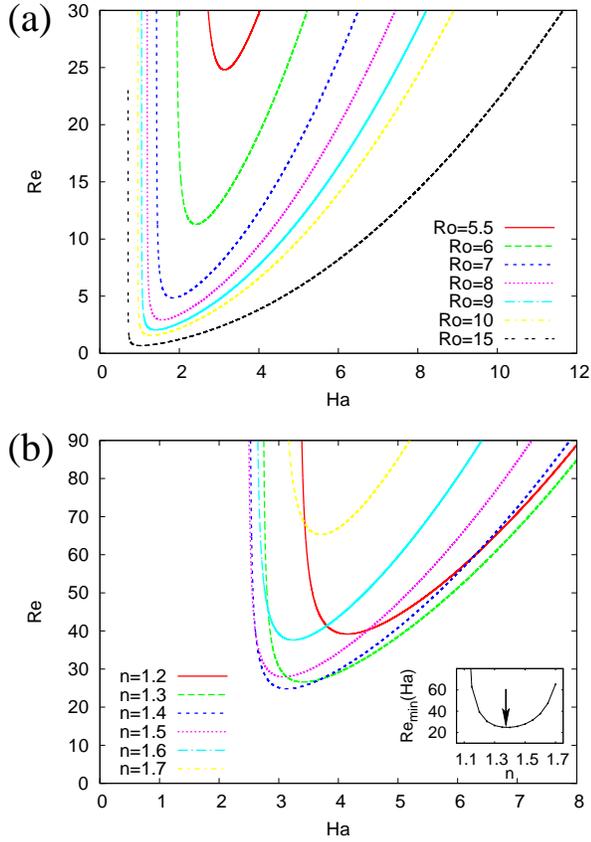}
    \end{center}
    \caption{Marginal curves for ${\rm Rb}=-1$. (a) Dependence on ${\rm Ro}$ for $n=1.4$. (b) 
    Dependence on $n$ for ${\rm Ro=5.5}$. The insert shows the dependence 
    of the minimum value  (with respect to ${\rm Ha}$) 
    of the critical ${\rm Re}$ on $n$. The arrow points to the 
    optimum $n\approx 1.35$ that leads to 
    the lowest critical ${\rm Re}$.}
    \label{fig1}
    \end{figure}

While for ${\rm Rb}=-1$ the only energy source of the instability
is the shear of the rotating flow, we move now in direction of ${\rm Rb}=0$ 
which corresponds to a constant
current density in the fluid, for which the 
kink-type TI \cite{SEIL1} is expected to occur. For the particular 
choice $n=1.2$, 
this transition is illustrated in Figure 2 where we have intensionally
chosen, for all ${\rm Rb}$, the same scales for ${\rm Re}$ and 
${\rm Ha}$.
For $\rm Rb=-0.6$ we observe the appearance of a crossing with the abscissa,
i.e. a point where the instability draws all its energy from the 
electrical current
instead of the shear. Actually, the lowest value where this can occur
is ${\rm Rb}=n^2/4-1=-0.64$ \cite{KS14}.  
 \begin{figure}
    \begin{center}
    \includegraphics[angle=0, width=0.45\textwidth]{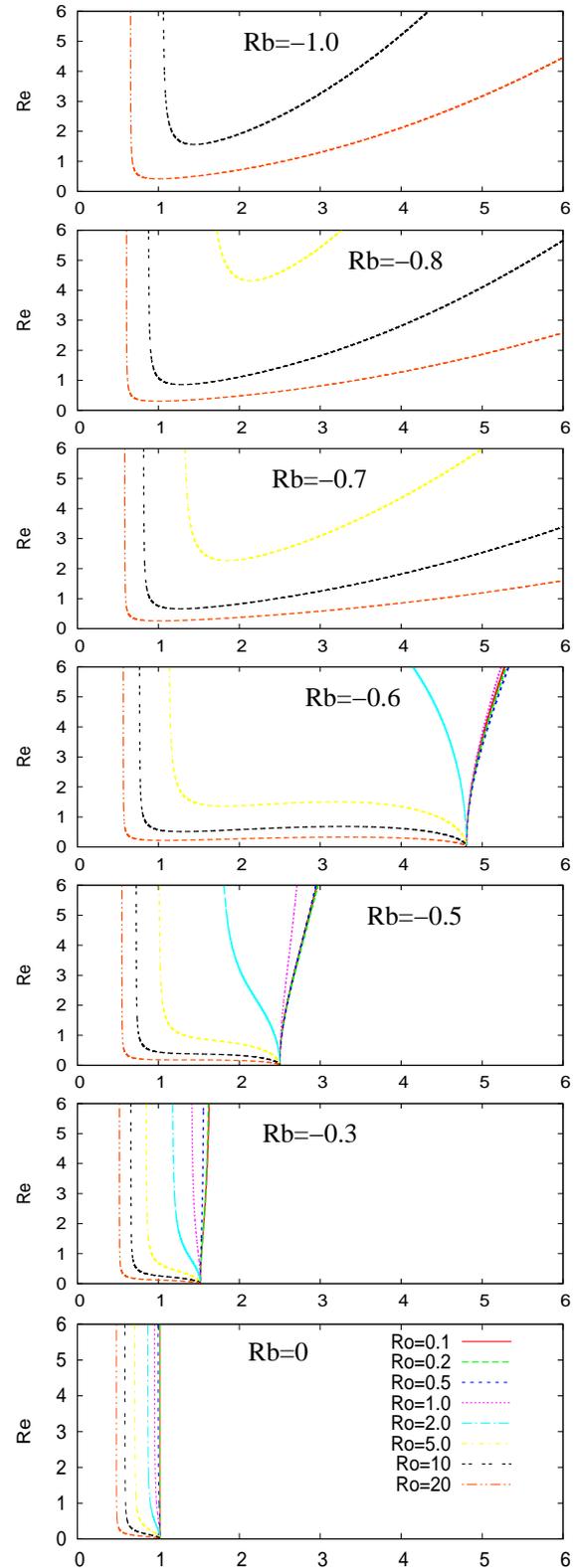}
    \end{center}
    \caption{Marginal curves for $n=1.2$ and various values of ${\rm Rb}$, in dependence 
    on ${\rm Ro}$. From top to bottom, the instability
    changes its character from a (magnetically triggered) shear-driven 
    instability to a (rotationally influenced) current-driven TI. 
    For $n=1.2$, TI 
    appears first for ${\rm Rb}=n^2/4-1=-0.64.$
    }
    \label{fig2}
    \end{figure}
\begin{figure}
    \begin{center}
    \includegraphics[angle=0, width=0.45\textwidth]{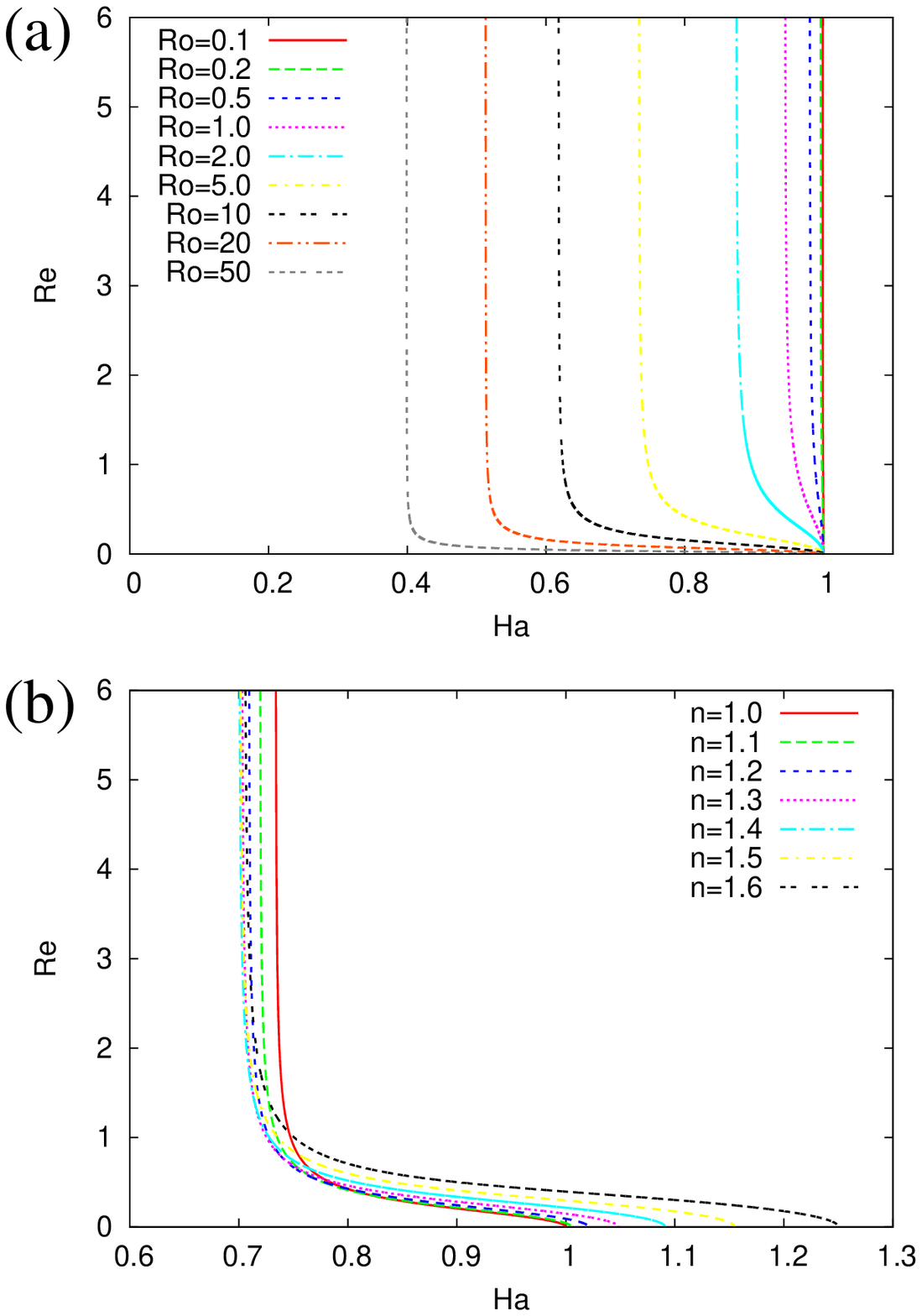}
    \end{center}
    \caption{Marginal curves for ${\rm Rb}=0$. (a) Dependence on ${\rm Ro}$ for $n=1.0$. (b) 
    Dependence on $n$ for ${\rm Ro=5}$.}
    \label{fig3}
    \end{figure}      
    
For $\rm Rb=0$ the instability is characterized in more detail in Figure 3. 
Very similar to the results of \cite{SUPER}, we observe in 
Figure 3a
that for $\rm Ro>0$ the curves move to the  left 
with increasing $\rm Re$
(i.e., the flow
{\it supports} the kink-type instability) and converge 
to well-defined values of $\rm Ha$ when $\rm Re$ goes to infinity.
The dependence on the wavenumber ratio $\alpha$ is quite
interesting. Figure 3b shows that the mode with $n=1$ (i.e. with $k_r=0$),
which is still dominant at $\rm Re=0$, is replaced by modes with higher 
values of $n$ for increasing $\rm Re$.
The limits of the critical $\rm Ha$ for $\rm Re=0$ and
$\rm Re\rightarrow \infty$ can be determined by setting to zero,
in Equation 2,
the nominator or denominator, respectively, 
which leads (for $\rm Rb=0$) to
\begin{eqnarray}
{\rm Ha}_{\rm Re=0}&=&1/\sqrt{n(2-n)}, \\
{\rm Ha}_{\rm Re \rightarrow \infty}
&=&\sqrt{    \frac{ ({\rm Ro}+1)+\sqrt{({\rm Ro}+1)}( {\rm Ro}+2)/n}   {{\rm Ro}^2+({\rm Ro}+1)(4-n^2) }          } .
\end{eqnarray}
In the limit $\rm Ro\rightarrow \infty$ the 
limit values of ${\rm Ha}$ converge slowly to zero according to
${\rm Ha}_{(\rm Re, Ro) \rightarrow \infty}\simeq  n^{-1/2} {\rm Ro}^{-1/4}$. 

\begin{figure}
    \begin{center}
    \includegraphics[angle=0, width=0.45\textwidth]{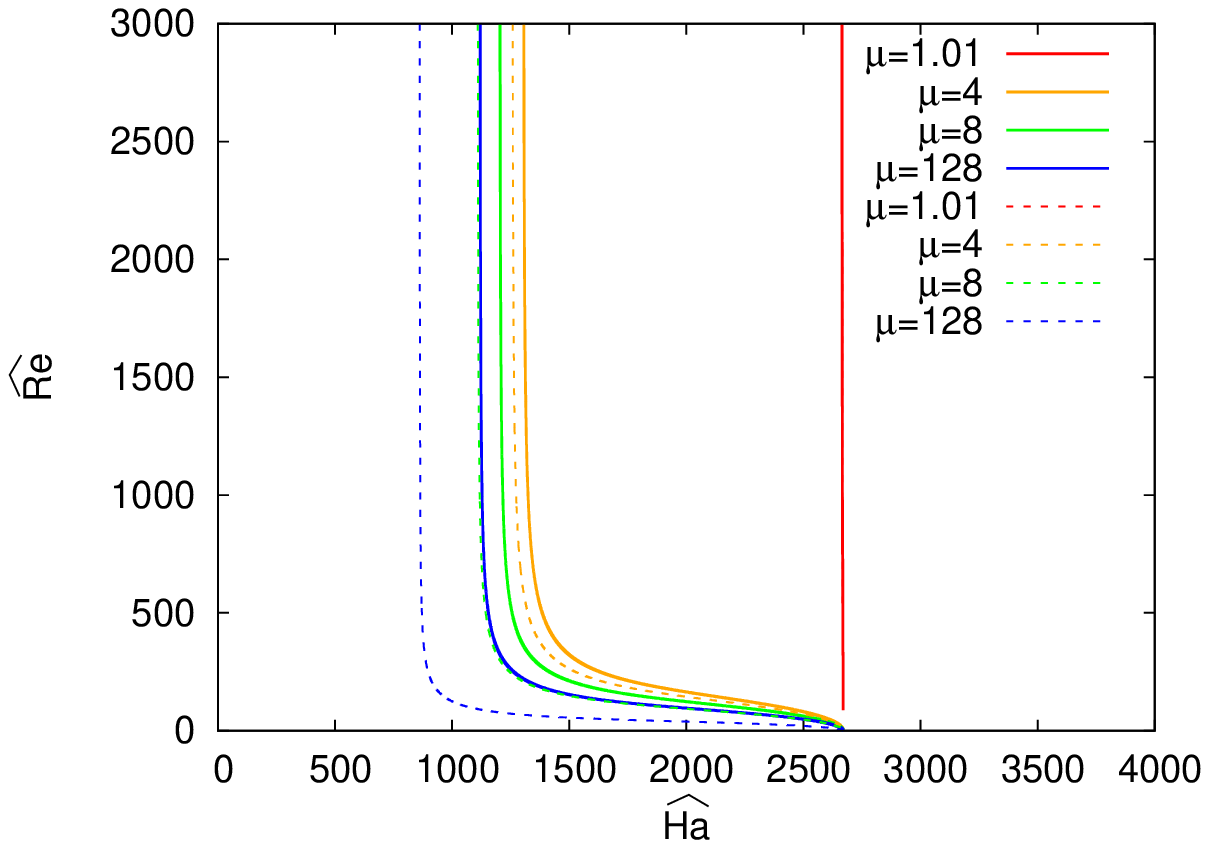}
    \end{center}
    \caption{Marginal curve for ${\rm Rb}=0$ and $n=1.41$, as scaled according 
    to 
    \cite{SUPER}. The full lines correspond to the translation of $\hat{\mu}$
    to $\rm Ro^{**}$, the dashed lines to $\rm Ro^{*}$.}
    \label{fig4}
    \end{figure}   
         
In the following, we compare our WKB results with recent 
findings \cite{SUPER} obtained for a TC flow with 
inner and outer radii $r_i$ and $r_o$ rotating with the angular
velocities $\Omega_i$ and $\Omega_o$, respectively. 
The corresponding ratios are defined as $\hat{\eta}=r_i/r_o$, and
$\hat{\mu}=\Omega_o/\Omega_i$. For this
TC configuration, the following modified 
definitions of the Reynolds and Hartmann 
number were used: ${\rm \widehat{Re}}=\Omega_o r_i(r_o-r_i)/\nu$,
${\rm \widehat{Ha}}=B_{\phi}(r_i) (r_i(r_o-r_i))^{1/2}/(\mu_0 \rho \nu \eta)^{1/2}$.
The non-trivial point is now how to translate the $\hat{\mu}$ 
of a TC flow, characterized by $\Omega(r)=a+b/r^2$,
to the $\rm Ro$ of a flow with $\Omega(r) \sim r^{2 Ro}$. 
An often used correspondence, based on 
equalizing the corresponding angular velocities at $r_i$ and $r_o$
\cite{MNRAS}, 
leads to 
\begin{equation}
\rm Ro^{*} \simeq - 1/2 log_{\hat{\eta}} \hat{\mu}
\end{equation}
while an alternative, more shear-oriented version leads to
\begin{equation}
{\rm {Ro}}^{**} \simeq \frac{1}{2} \frac{(1+\hat{\eta})(\hat{\mu}-1)}{(1-\hat{\eta})(\hat{\mu}+1)} \; .
\end{equation}
Actually, for comparably small (positive or negative) values
of $\rm Ro$, the differences are not very significant,
but they increase for steeper profiles.
This is a key point for the adequateness of
TC flows to ``emulate'' steep power function flows.
In \cite{SUPER}, the destabilizing effect of positive shear 
had been studied for TC flows (with
$\rm Rb=0$ only), both for a wide gap with $\hat{\eta}=0.5$ as well 
as a narrow gap 
with $\hat{\eta}=0.95$. In either case, for 
large values of $\hat{\mu}$,
the critical $\rm Ha$ converged to some non-zero constant, 
which is not compatible with the translation to $\rm Ro^{*}$
since the latter should lead to a zero critical $\rm Ha$ 
(according to ${\rm Ha}_{(\rm Re, Ro) \rightarrow \infty}\simeq  
n^{-1/2} {\rm Ro}^{-1/4}$, see above). It turns out that 
the translation to $\rm Ro^{**}$ is physically more adequate.

With the reasonable choice $k_z=k_r=\pi/(r_i-r_o)$ we
obtain the translations
 ${\rm \widehat{Re}}=\pi^2 2^{5/2} \hat{\mu} \hat{\eta}/((1+\hat{\mu})(1-\hat{\eta})) \; {\rm {Re}}$
 and 
$\rm \widehat{Ha}=\pi^2 (1+\hat{\eta})^2/((2 \hat{\eta})^{1/2} (1-\hat{\eta})^{3/2}) \; \rm{Ha}$.
For $\hat{\eta}=0.95$ this amounts to 
${\rm \widehat{Re}}= 1061/(1+1/\hat{\mu}) {\rm {Re}}$ 
and
${\rm \widehat{Ha}}=2435 {\rm {Ha}}$. 
Figure 4 shows the corresponding WKB results, both for 
assuming a translation to $\rm Ro^{*}$ (dashed lines) and 
to $\rm Ro^{**}$ (full lines).
For ${\rm \widehat{Re}}=0$ our result ${\rm \widehat{Ha}}=2670$
agrees reasonably well with the exact value 
${\rm \widehat{Ha}}=3060$ of
the modal stability analysis \cite{SUPER}. What is 
more, the typical bend of the marginal curve 
to the left for increasing ${\rm \widehat{Re}}$,
and the limit values of ${\rm \widehat{Ha}}$ for large ${\rm \widehat{Re}}$,
are also confirmed.
Yet, subtle differences show up for the two ways of translation: 
the use of $\rm Ro^{**}$
confirms the existence of a finite limit value for the critical
${\rm \widehat{Ha}}$, as typical for TC flows, while the use of 
$\rm Ro^{*}$ would ultimately lead to a zero limit value.

This encouraging consistency of the local approximation and the 
modal stability analysis, evidenced for $\rm Rb=0$, 
brings us back to the 
point whether, for $\rm Rb=-1$, the ULL can be 
confirmed in a TC experiment. 
Assuming $\rm Ro^{**}$ as more physical than 
$\rm Ro^{*}$, in the limit ${\hat{\mu} \rightarrow \infty}$
we obtain 
$\rm Ro^{**}_{\hat{\mu} \rightarrow \infty}=1/2 (1+\hat{\eta})/(1-\hat{\eta})$. 
This means, in turn, that 
to emulate some ${\rm {Ro}}$ in a TC-flow, 
$\hat{\eta}$ has to fulfill the relation 
$\hat{\eta}=(2 {\rm {Ro}} -1)/(2 {\rm {Ro}} +1)$. With view 
on the ULL, this implies that for ${\rm {Ro}}=6$, say, 
a minimum value of $\hat{\eta}=11/13=0.846$ is needed. For TC-flows with 
wider gaps, such as $\hat{\eta}=1/2$ the necessary 
shear could simply not be realized.

What are, then, the prospects for a corresponding experiment?
Evidently, we need a rather narrow gap flow. Let us stick, for a first 
estimate, to the safe value $\hat{\eta}=0.95$, and take the 
typical values $\rm Ro=6$, $\rm Ha=2$ and
$\rm Re=12$ as read off from Figure 1a. This translates to 
$\hat{\mu}=1.89$, ${\rm \widehat{Re}}=8324$ and 
${\rm \widehat{Ha}}=4870$. For a prospective 
TC experiment with Na at 150$^\circ$C, 
with $\rho=910$ kg/m$^3$, $\nu=5.94\times 10^{-7}$ m$^2$/s, 
$\sigma=9 \times 10^6$ S/m, 
and an outer diameter of $r_o=0.25$ m, this would amount to 
a rather moderate 
rotation frequency of $\Omega_o/(2 \pi)=0.26$ Hz, yet a 
huge magnetic field $B_{\phi}(r_i)=0.69$ T that requires 
a central current of $I=8.6 \times 10^5$ A. Exhausting the
shear resources, by choosing ${\hat{\mu}} \rightarrow \infty$ and
$\hat{\eta}=0.85\approx 11/13$, those values would drop
to ${\rm \widehat{Re}}=3796$,  ${\rm \widehat{Ha}}=892$
or, physically, to $\Omega_o/(2 \pi)=0.044$ Hz, 
$B_{\phi}(r_i)=77$ mT, $I=8.2 \times 10^4$ A.
Any real TC experiment, however, would need 
more detailed simulations with a 1D marginal stability 
code to confirm and optimize the parameters. 

This work was supported by German Helmholtz Association in frame 
of the Helmholtz Alliance LIMTECH. F.S. gratefully acknowledges 
fruitful discussions with G\"unther R\"udiger.


\begin{thebibliography}{30}


\bibitem{RAYLEIGH}
Lord Rayleigh, 
Proc. R. Soc. London A \textbf{93}, 148 (1917).

\bibitem{VELI}
E.P. Velikhov, JETP \textbf{9}, 995 (1959); 
S.A. Balbus, J.F. Hawley,
Astrophys. J. \textbf{376}, 214 (1991)

\bibitem{LIU2006}
W. Liu, J. Goodman, H. Ji,
Astrophys. J. \textbf{643}, 306 (2006)


\bibitem{BH08}
S.A. Balbus, P. Henri,
Astrophys. J.  \textbf{674}, 408 (2008)



\bibitem{HR95}
R. Hollerbach, G. R\"udiger,
Phys. Rev. Lett. \textbf{95}, 124501 (2005)

\bibitem{P11}
J. Priede,
Phys. Rev. E \textbf{84}, 066314 (2011)


\bibitem{LIU}
W. Liu, J. Goodman, I. Herron, H. Ji,
Phys. Rev. E \textbf{74}, 056302 (2006)

\bibitem{KS14}
O.N. Kirillov, F. Stefani, Y. Fukumoto,
J. Fluid Mech. \textbf{760}, 591 (2014)

\bibitem{KS13}
O.N. Kirillov, F. Stefani,
Phys. Rev. Lett. \textbf{111}, 061103 (2013)




\bibitem{TEELUCK}
R. Hollerbach, V. Teeluck, G. R\"udiger,
Phys. Rev. Lett. \textbf{104}, 044502 (2010)

\bibitem{APJ12}
O.N. Kirillov, F. Stefani, Y. Fukumoto,
Astrophys. J. \textbf{756}, 83 (2012)

\bibitem{PRE}
F. Stefani et al., Phys. Rev. Lett. \textbf{97}, 184502 (2006); 
F. Stefani et al., Phys. Rev. E. \textbf{80}, 066303 (2009)

\bibitem{SEIL2}
M. Seilmayer et al., Phys. Rev. Lett. \textbf{113}, 024505 (2014)

\bibitem{TSUKAHARA}
T. Tsukahara, N. Tillmark, P.H. Alfredsson,
J. Fluid Mech. \textbf{648}, 5 (2010)


\bibitem{PAME} K.P. Parfrey, K. Menou, 
Astrophys. J. Lett. \textbf{667}, L207 (2007)

\bibitem{CHA} P. Charbonneau, Liv. Rev. Sol. Phys. \textbf{7}, 3 (2010)

 
 
\bibitem{RUEPERS}
G. R\"udiger, personal communication

\bibitem{SEIL1}
G. R\"udiger, M. Schultz, 
Astron. Nachr. {\textbf 331},
121 (2010);
M. Seilmayer et al., Phys. Rev. Lett. \textbf{108}, 244501 (2012)

\bibitem{SUPER}
G. R\"udiger et al., Phys. Fluids, subm. (2015); arXiv:1505.05320

\bibitem{MNRAS} G. R\"udiger et al., Mon. Not. R. Astron. Soc, {\textbf 438},
271 (2014)




\end{thebibliography}
\end{document}